\documentstyle[12pt]{article}

\def\xpx{x\partial_x}
                                                                 
\begin{document}
                                                                 
 \begin{titlepage}
January  1995
\begin{center}                                                   
{\LARGE\bf Polymer-Chain Adsorption Transition}
\medskip
{\LARGE\bf at a Cylindrical Boundary}

\vspace{2.0cm}                                                   
 \large{
S. Boettcher\\
Department of Physics, \\
Brookhaven National Laboratory, Upton, NY 11973, USA
\\ }

\end{center}
\vspace{4.0cm}
\abstract{
In a recent letter, a simple method was proposed to generate solvable
models that predict the critical properties of statistical systems
in hyperspherical geometries. To that end, it was shown how to
reduce a random walk in $D$ dimensions to an anisotropic one-dimensional 
random walk on concentric hyperspheres. Here, I
construct such a random walk to model the adsorption-desorption
transition of polymer chains growing near an attractive cylindrical
boundary such as that of a cell membrane. I find that the fraction
of adsorbed monomers on the boundary vanishes exponentially when
the adsorption energy decreases towards its critical value. When
the adsorption energy rises beyond a certain value above the
critical point whose scale is set by the radius of the cell, the
adsorption fraction exhibits a crossover to a linear increase
characteristic to polymers growing near planar boundaries. 
}
\vfill
\end{titlepage}

\section{Introduction}

The study of polymers is rightfully popular with experimentalist
and theorist alike.\cite{AIP,DesCJ} Their diversity and practical
importance in biology and chemistry makes experimental investigations
rewarding. The variety and complexity of polymeric systems arising
from a few simple building blocks inspires much theoretical and
numerical work using techniques such as scaling,\cite{DeGennes}
renormalization group,\cite{Freed} and Monte Carlo calculations.\cite{Binder} 
While we are far from a solvable first-principles
model of polymer systems resembling reality, the conception
of simplified statistical models might allow us to capture elementary
aspects of some of the critical phenomena exhibited by real polymers.
Thus, even simple models of polymers, like any solvable statistical
model with nontrivial critical behavior, warrants attention because
of the insight granted into the interplay of fundamental properties
such as the range of forces, the symmetries of the system, and
geometrical constraints.\cite{Baxter}

The simplest polymer system is that of an unbranched chain of
monomers embedded in some volume. While practically difficult to
obtain, it is simple to model this system by a self-avoiding
random walk (SAW).\cite{FloMo} The properties of random walk models
can easily be studied numerically, and in some cases even
analytically.\cite{Sokal} In this paper, I present the study of a
polymer growing in the neighborhood of an attractive boundary. The
properties of this polymer can be derived analytically when modeled
as a directed SAW on a lattice that only allows for rectangular
turns. Earlier investigations have focused on a polymer growing
in the neighborhood of a planar boundary.\cite{Priv} Here, I use a
recently proposed model for random walks on a hyperspherical
lattice to extend the theory to $D$-dimensional hyperspherical
boundaries.\cite{BeBoM} I then give a detailed discussion of
the solution for the adsorption fraction of an infinitely long 
polymer, $P(\kappa)$, at an attractive cylindrical boundary with a 
potential $\kappa$.\cite{BoMo} Here, the adsorption fraction is
nonvanishing  only if the attractive potential on the boundary is larger
than a critical value $\kappa_c$. I show that the critical properties of the
adsorption fraction in this case are profoundly different to
those found near planar boundaries. While the adsorption fraction scales
linearly at a planar boundary, I find for a cylinder of radius  
$m\geq 0$ in monomer units and for $\Delta
\kappa\equiv \kappa-\kappa_c \to 0_+$, that the asymptotic behavior of
$P(\kappa)$ is given by
\begin{equation}
P(\kappa)\sim {4\over 81} {e^{-{8\over 9(m+1) \Delta \kappa}}\over
(m+1)\Delta \kappa^2}. 
\end{equation}
If the radius $m$ of the boundary is large compared to the length of a
monomer, one would expect to recover the scaling behavior that is
characteristic of a planar boundary except for the immediate
neighborhood of the critical point. In fact, I find for $m\ll 1$
that there is a crossover between linear and exponential scaling when
$m\Delta\kappa\sim 1$. That is, if $m\Delta\kappa\ll 1$, then Eq.~(1)
holds, while linear scaling is obtained as soon as $m\Delta\kappa\gg 1$.

The simplicity of the dynamics in this new random walk model, and
the nontrivial phenomena obtained from it, raise interesting questions
regrading the universal properties of this lattice. According to 
universality,\cite{PliBer} at the critical transition only a few 
fundamental properties of the system determine 
its behavior. In this model, the critical behavior arises from the
balance between a short-range attractive potential and the spatial entropy.
I argue that the entropy at the
critical point is sufficiently well represented by a hyperspherical lattice.
It has been shown that such a lattice reproduces all the universal features
expected of lattices.\cite{BeBoM} The critical behavior obtained on this
lattice for rotationally symmetric systems should therefore reflect the
universal critical behavior of the system. The advantage of random walks 
on hyperspheres is to describe the critical behavior in a minimal 
and tractable way in comparison, for example, to a far more structured 
hypercubic lattice.

In the following section, I generalize the theory for a polymer
growing at an attractive boundary in a $D$-dimensional hyperspherical
geometry. In Sec.~III, I discuss the theory for the special cases
$D=1,~2$, and 3. In Sec.~IV, I analyze in detail the solution
obtained for $D=2$, the case of an attractive cylindrical boundary.
I evaluate the solution asymptotically near the critical point to
derive Eq.~(1). Finally, in Sec.~V I discuss some implications of my findings.

\newpage
\section{Directed Walk Model for the Polymer Adsorption Transition}

In this paper I want to describe the growth of a single-stranded polymer 
at an attractive hyperspherical boundary.  The 
boundary is considered to be impenetrable. The first monomer in this 
polymer chain is grafted to the boundary. New monomers are
added to the end of the existing polymer chain in a random fashion.
The addition of each monomer obtains a fugacity of $z$, while the
addition of a monomer on the boundary yields an energy gain for the
system of $\kappa\geq 1$. For simplicity, I want the polymer to be stretched
out (directed) along the boundary to avoid self-interaction and
excluded-volume effects.

Polymers in solution at an attractive boundary undergo an 
adsorption-desorption transition at a critical value of the attractive
potential, $\kappa_c$.\cite{Taka} From the theory for a single-stranded 
polymer it is found that $P(\kappa)$, the fraction of adsorbed monomers
in the limit of an infinitely long polymer, attains
a finite value only for $\kappa$ larger than $\kappa_c$.\cite{Priv}
Here, I am interested in the critical properties of the adsorption
fraction for a polymer as a function of the curvature of the
attractive boundary.

The growth process of a polymer can be modeled as a directed,
$D+1$-dimensional random walk in a $D$-dimensional hyperspherical 
geometry. (The extra dimension refers to the time coordinate of the walk
which corresponds to the length $L$ of the polymer extending alongside the
$D$-dimensional hyperspherical boundary.) I choose for this 
random walk to occur on a lattice consisting of a set of concentric
hyperspherical surfaces, $S_n$, equally spaced in units of a monomer 
length. The innermost surface $S_m,~m\geq 0,$ representing the attractive 
boundary, has an integer radius of $m$ in monomer units, the next surface 
$S_{m+1}$ has a radius of $m+1$, and so on. The (virtual) surface 
area of the $n$th surface is 
\begin{equation}
S_n={2\pi^{D/2}\over\Gamma(D/2)}\,n^{D-1}.
\end{equation}
The walker resides before each step in some region $n\geq m$ between
the surface $S_n$ and the surface $S_{n+1}$. In general, the statistical
weights for proceeding are dependent on the location of the
walker and the direction of the next step due to the anisotropy
of the spherical geometry. A step parallel to the boundary to stay
in region $n$ is taken with some weight $P_{stay}(n)$, while the
weights for the walker to either move outward to region $n+1$ or
move inward to region $n-1$ may be given by $P_{out}(n)$ and
$P_{in}(n)$, respectively. For the case of a cylindrical boundary,
$D=2$, such a walk is shown in Fig.~1.

For a growing polymer or a random walker above a convex boundary like that 
of a $D$-dimensional hypersphere, there is on each step always more 
``open space'' (i. e. accessible states) available in proceeding outward 
than in proceeding inward. This anisotropy is more pronounced for stronger
curved boundaries, i. e. increasing $D$. A walker anywhere in region
$n$ has for an inbound step a number of target states available
that is proportional to the virtual surface  area $S_n$ that is
traversed. For an outbound step, the number of states available
is proportion to the area $S_{n+1}$. Using Eq.~(2), this effect
can be accounted for by assigning\footnote{Overall factors are chosen
to conform to Ref.~\cite{Priv} when $D=1$ or $m\to\infty$.} for all $n>m$
\begin{eqnarray}
P_{stay}(n)&\equiv& 1,\cr 
\noalign{\medskip}
P_{in}(n)&=& {2 S_n\over S_n+S_{n+1}}={2 n^{D-1}\over
{n^{D-1}+(n+1)^{D-1}}}, \cr
\noalign{\medskip}
P_{out}(n)&=&{2 S_{n+1}\over S_n+S_{n+1}}={2 (n+1)^{D-1}\over
{n^{D-1}+(n+1)^{D-1}}},
\end{eqnarray}
whereas on the boundary $S_m$, I choose
\begin{equation}
P_{stay}(m)=1,\quad 
P_{in}(m)=0, \quad
P_{out}(m)=1.
\end{equation}
Note, that the walk is prohibited from reaching inward from region 
$m$ by setting $P_{in}(m)=0$. The walk starts in region $m$.
Every step is further weighted with a factor $z$, and each parallel
step inside the region $m$ acquires an additional factor of $\kappa$.

A walk with $L>0$ parallel steps has reached $L$ levels,
$\{h_i\}_{i=1}^{L+1},~h_i\geq m$, above or at the boundary. I want
to restrict these walks such that $\vert h_{i+1}-h_i \vert \leq
1,~0\leq i\leq L$. This restriction on the directed walk, in the spirit
of a restricted solid-on-solid model (RSOS), has no impact on the
critical behavior of the system for an ensemble average of walks of all
length $L$. But this restriction implies that each random step that leads 
into a radial direction must be followed by a {\sl deterministic} parallel 
step. Thus, while the walk in Fig.~1 consists of $N=26$ links, only $L=14$ 
random choices of levels $h_i,~1\leq i\leq L$, were require.

A transfer matrix $T_{h_{i+1},h_i}$ that describes the transition of the
walker from the $i$th to the $(i+1)$st level is then given by
\begin{equation}
T_{j,i}=z^{\vert j-i\vert} \kappa^{\delta_{m,j}}\left [P_{stay}(i)
\delta_{j,i} + P_{out}(i) \delta_{j-1,i} + P_{in}(i) \delta_{j+1,i}
\right ].
\end{equation}
The transfer matrix $T$ is in general asymmetric because of the
anisotropy of the hyperspherical geometry. It is symmetric only
for the case of a planar geometry, $D=1$, where $P_{in}\equiv
P_{out}$. The total statistical weight of a certain walk
configuration is then given by
\begin{equation}
z^L \delta_{m,h_0} T_{h_1,h_0} T_{h_2,h_1}\ldots T_{h_L,h_{L-1}},
\end{equation}
such that the partition function $Z_L$ for all walks extending $L$
parallel steps away from the starting point is
\begin{equation}
Z_L=z^L {\vec b}^{(t)} T^L {\vec e},
\end{equation}
where ${\vec b}^{(t)}$ and ${\vec e}$ are vectors accounting for
beginning and end effects. The total partition function for walks
of all length, $Z=\sum_{L=1}^{\infty} Z_L$, then evaluates to
\begin{equation}
Z(z,\kappa)=z {\vec b}^{(t)} T (1-zT)^{-1} {\vec e}.
\end{equation}
For any given $\kappa$, if $\lambda_{max}(\kappa,z)$ is the largest 
eigenvalue of $T$, then $Z$
diverges for $z\nearrow z_{\infty}(\kappa)$, where $z_{\infty}(\kappa)$
is defined by
\begin{equation}
1=z_{\infty}(\kappa)\lambda_{max}(\kappa,z_{\infty}(\kappa)).
\end{equation}

The average length of a walk is usually defined to be 
\begin{equation}
\langle N(z,\kappa)\rangle=z \partial_z \ln{Z(z,\kappa)},
\end{equation}
whereas the average number of steps taken on the boundary can be
obtained through
\begin{equation}
\langle N_s(z,\kappa)\rangle= \kappa \partial_\kappa
\ln{Z(z,\kappa)}. \end{equation}
Both, $\langle N\rangle$ and $\langle N_s\rangle$, diverge like some
power of $1/[z_{\infty}(\kappa)-z]$ for $z\nearrow z_{\infty}(\kappa)$.
Thus, for $z\nearrow z_{\infty}(\kappa)$, the limit is obtained where the 
average polymer is infinitely long. At the same time, $\langle N_s\rangle$ 
refers to the number of monomers which are adsorbed on the boundary as a
function of the attractive boundary potential $\kappa$. The fraction
of adsorbed monomers $P(\kappa)$ is given by
\begin{equation}
P(\kappa)=\lim_{z\nearrow z_{\infty}(\kappa)} {\langle
N_s(z,\kappa)\rangle \over \langle N(z,\kappa)\rangle }=-{\kappa
\over z_{\infty}(\kappa)}{d z_{\infty}(\kappa) \over d\kappa}.
\end{equation}
Thus, $z_{\infty}(\kappa)$ marks a line in the $(\kappa,z)$-plane
for which $P(\kappa)$ is defined. 

While this random walk model is {\sl in its details} only a
crude description of a polymer growing in a continuum, it can be
expected that the critical properties of the polymer system in the
infinite chain limit is well approximated by such a model. The
critical properties of the polymer system arise from the balance
between the entropy of all possible polymer (or walk) configurations
in the space above the boundary and the energy gained in the
attractive potential on the boundary. One might anticipate that
the critical properties of this system will vary in an interesting
way as a function of the curvature of the boundary. For instance,
the more the boundary is curved, the larger is the space available
for walk configurations away from the boundary and, therefore, the larger
is the entropy. I will show that for a curved boundary the critical 
transition is substantially weaker than in the case of a planar boundary.

\section{The Adsorption Problem for Planar, Cylindrical, and
Spherical Boundaries}

In this section, I use generating function techniques to derive a
differential equation that has to be solved to obtain the spectrum of
the transfer matrix $T$. First, I discuss the differential operator for
arbitrary $D$. Then, I examine the differential equation for the special
cases of $D=1,~2$, and 3.

To determine the spectrum $\lambda$ of the transfer matrix $T$, I insert
the weights in Eqs.~(3-4) into Eq.~(5), leading to the eigenvalue problem
\begin{eqnarray}
\lambda g_n&=&\sum_{i=m+1}^{\infty} T_{n,i}\, g_i \cr
\noalign{\bigskip}
&=&\cases{
g_n+z {2n^{D-1}\over {(n-1)^{D-1}+n^{D-1}}} g_{n-1}+z {2(n+1)^{D-
1}\over {(n+1)^{D-1}+(n+2)^{D-1}}} g_{n+1},& $n\geq m+2$;\cr
\noalign{\medskip}
g_{m+1}+z g_m+z {2(m+2)^{D-1}\over {(m+2)^{D-1}+(m+3)^{D-1}}}
g_{m+2},& $n=m+1$;\cr
\noalign{\medskip}
\kappa g_m+\kappa z {2(m+1)^{D-1}\over {(m+1)^{D-1}+(m+2)^{D-1}}}
g_{m+1},& $n=m$.\cr}
\end{eqnarray}
This system always has a continuous spectrum, independent of $\kappa$,
but the spectrum contains bound states only for a certain range of
$\kappa$. If for a neighborhood of a value of $\kappa$ the largest
eigenvalue $\lambda_{max}$ is given by the upper bound on the continuous
spectrum, then the adsorption fraction vanishes due to Eq.~(12), since
$\lambda_{max}$ -- and thus $z_\infty$ -- does not vary with $\kappa$.
On the other hand, if a bound state exists, its eigenvalue usually
varies as a function of $\kappa$ and is larger than the continuous
spectrum, which leads to a nonvanishing adsorption fraction.
Thus, the existence of bound states will prove to be the criterion for the
appearance of an adsorbed phase for the polymer. A bound state has to 
satisfy a condition ensuring that the likelihood of finding the walker 
in far-out regions $n\to\infty$ is diminishing sufficiently fast, i. e.
\begin{equation}
g_n\to 0 {\rm~for~} n\to\infty.
\end{equation}
It is convenient to define
\begin{equation}
g_n=\cases{ \left[ n^{D-1}+(n+1)^{D-1}\right] h_n,& $n>m$;\cr
\noalign{\medskip}
2(m+1)^{D-1} h_m,& $n=m$.\cr}
\end{equation}
Then, Eqs.~(13) reduce to
\begin{equation}
0=(1-\lambda)\left[n^{D-1}+(n+1)^{D-1}\right] h_n+2z n^{D-1}
h_{n-1}+2z (n+1)^{D-1} h_{n+1},\quad n>m,
\end{equation}
and
\begin{equation}
0=(\kappa-\lambda) h_m + \kappa z h_{m+1}.
\end{equation}
To simplify the analysis of this problem, I define the generating
functions 
\begin{equation}
G(x)=\sum_{n=m}^{\infty} g_n x^n,
\end{equation}
and
\begin{equation}
H(x)=\sum_{n=m}^{\infty} h_n x^n.
\end{equation}
Using the identity
\begin{equation}
\sum_n n^i x^n h_n = (\xpx)^i \sum_n x^n h_n,
\end{equation}
$G(x)$ can be formally obtained from $H(x)$ via
\begin{equation}
G(x)=\left[ (\xpx)^{D-1}+{1\over x}(\xpx)^{D-1}x\right] H(x) -
\left[(\xpx)^{D-1}-{1\over x}(\xpx)^{D-1}x\right] x^m h_m,
\end{equation}
when $D$ is a positive integer. Note that $(\xpx)^{D-1} x^m=m^{D-
1} x^m$ for $m>0$ or $D>1$, but that $(\xpx)^{D-1} x^m=1$ for
$D=1$ and $m=0$.
The eigenvalue problem can be converted from a difference equation
for $h_n$ into a differential equation for $H(x)$. The condition
in Eq.~(14) on $g_n$ translates into an analyticity condition on
$G(x)$: The exponentially growing solutions for $g_n$ in Eq.~(13) are 
averted, if and only if $G(x)$ has no singularities for $|x|<1$. Thus, 
one obtains bound states exactly when singularities of $G(x)$ appear inside
of the unit circle with eigenvalues that are determined by a
condition that cancels these singularities.

The differential form of the eigenvalue problem is obtained by
multiplying Eq.~(16) with $x^n$ and summing from $n=m+1$ to
$n=\infty$:
\begin{eqnarray}
0&=&(1-\lambda)\left[\sum_{n=m+1}^{\infty} n^{D-1} x^n h_n
+\sum_{n=m+1}^{\infty} (n+1)^{D-1} x^n h_n\right] \cr
\noalign{\medskip}
&&\quad +2z\sum_{n=m+1}^{\infty} n^{D-1} x^n h_{n-1} +
2z\sum_{n=m+1}^{\infty} (n+1)^{D-1} x^n h_{n+1}.
\end{eqnarray}
After shifting indices and applying the identity in Eq.~(20), one
gets
\begin{eqnarray}
0&=&(1-\lambda)\left[(\xpx)^{D-1}+{1\over x}(\xpx)^{D-1}x\right]
\sum_{n=m+1}^{\infty} x^n h_n\cr
\noalign{\medskip}
&&\quad +2z(\xpx)^{D-1}x \sum_{n=m}^{\infty} x^n h_n + 2z{1\over
x}(\xpx)^{D-1} \sum_{n=m+2}^{\infty} x^n h_n.
\end{eqnarray}
Completing the sums by adding and subtracting the missing terms
in the definition of $H(x)$ in Eq.~(19), one obtains
\begin{eqnarray}
0&=&(1-\lambda)\left[(\xpx)^{D-1}+{1\over x}(\xpx)^{D-1}x\right]
\left[H(x)-x^m h_m\right]\cr
\noalign{\medskip}
&&\quad +2z(\xpx)^{D-1}x H(x) + 2z{1\over x}(\xpx)^{D-1} \left[H(x)-
x^m h_m -x^{m+1} h_{m+1}\right].
\end{eqnarray}
Eliminating $h_{m+1}$ by applying the boundary condition in Eq.~(17),
and separating off the inhomogeneous part gives
\begin{eqnarray}
\left\{(1-\lambda)\left[(\xpx)^{D-1}+{1\over x}(\xpx)^{D-1}x\right] 
+2z(\xpx)^{D-1}x + 2z{1\over x}(\xpx)^{D-1}\right\} H(x)=&\cr
\noalign{\medskip}
\left\{(1-\lambda)\left[(\xpx)^{D-1}+{1\over x}(\xpx)^{D-
1}x\right] + 2z{1\over x}(\xpx)^{D-1} +2\left({\lambda\over\kappa}-
1\right) {1\over x} (\xpx)^{D-1}x \right\} x^m h_m. &
\end{eqnarray}
To further simplify the presentation, I multiply both sides by
$x/(2z)$ and abriviate
\begin{eqnarray}
\epsilon&=&{2z\over \lambda-1},\cr
\noalign{\medskip}
\gamma&=&{1\over\epsilon} \left( 1-\sqrt{1-\epsilon^2}\right),\cr
\noalign{\medskip}
Q(x)&=&\sqrt{(x-\gamma)(x-1/\gamma)},
\end{eqnarray}
to obtain
\begin{eqnarray}
&&\left\{Q(x)^2 (\xpx)^{D-1} + Q(x) Q'(x) \left[(\xpx)^{D-1}
,x\right]_{-}\right\} H(x)\cr
\noalign{\medskip}
&&\quad=\left\{\left(1-{x\over\epsilon}\right)(\xpx)^{D-1} +
\left[{1\over \kappa} \left({2\over\epsilon}+{1\over z}\right)-
\left({1\over\epsilon}+{1\over z}\right)\right](\xpx)^{D-1}x \right\}
x^m h_m,
\end{eqnarray}
where $[.,.]_{-}$ is the usual commutator. Clearly, the operators
in the inhomogeneous part are easy to evaluate but remain as a
shorthand notation to avoid ambiguities for the case $D=1,~m=0$.
To proceed further in the analysis of this differential equation,
it is necessary to choose a specific positive integer $D$.
For a given $D$, one obtains an inhomogeneous differential equation
of order $D-1$. The solution of this differential equation is
uniquely determined through conditions imposed at the origin on
$H(x)$ and its derivatives, using the definition of $H(x)$ in
Eq.~(19). 

It is obvious from the differential equation in (27) that $H(x)$,
and therefore $G(x)$, generically has singularities at least at
$x=\gamma$ and $x=1/\gamma$. If $\gamma$ is real and positive,
then one of these singularities must be located inside of or on
the edge of the disk $|x|<1$. Say, that $\gamma<1$. Then, to avoid
growing solutions of the form $g_n\propto\gamma^{-n}$, a discrete
eigenvalue $\lambda$ exists which is determined through the condition 
\begin{equation}
\lim_{x\to\gamma} |G(x)|<\infty.
\end{equation}

For specific values of $D$, the left-hand side of Eq.~(27) can be
simplified further. For instance, for the planar case, $D=1$,
Eq.~(27) is especially simple, because all differential operators
disappear, the commutator vanishes, and the relation reduces to an
algebraic equation
\begin{equation}
G(x)=2H(x)={2\over Q(x)^2} \left\{1+ \left({1\over \kappa}-1\right)
\left({2\over\epsilon}+{1\over z}\right)x\right\} x^m h_m.
\end{equation}
Applying the condition in Eq.~(28) yields the relation for the
eigenvalues
\begin{equation}
0=1+ \left({1\over \kappa}-1\right) \left({2\over\epsilon} + {1\over
z}\right)\gamma
\end{equation}
that was discussed in Ref.~\cite{Priv}. Note that despite the
explicit appearance of the curvature radius $m$ in Eq.~(29), the
eigenvalue relation in Eq.~(30) is independent of $m$ as it should
be for the planar case. It is a necessary condition on the eigenvalue
relation for any $D$ that it reduces to Eq.~(30) for a hypersphere of
infinite radius, $m\to\infty$.

For $D=2$, the case of a polymer growing alongside an attractive
cylindrical boundary of radius $m$, one gets $[\xpx,x]_-=x$, and
Eq.~(27) reads
\begin{eqnarray}
&&Q(x)\left\{Q(x) \xpx + [\xpx Q(x)]\right\} H(x)\cr
\noalign{\bigskip}
&&\quad=Q(x)\xpx \left\{Q(x) H(x)\right\}\cr
\noalign{\bigskip}
&&\quad=\left\{\left(1-{x\over\epsilon}\right)m +\left[ {1\over
\kappa} \left({2\over\epsilon}+{1\over z}\right)-\left(
{1\over\epsilon}+{1\over z}\right)\right](m+1)x \right\} x^m h_m.
\end{eqnarray}
Requiring that $\lim_{x\to 0} x^{-m}H(x)=h_m$, determines the solution,
\begin{equation}
H(x)={m h_m\over Q(x)} \int_0^x dt {t^{m-1}\over Q(t)} \left(1+A
t\right),
\end{equation}
where I defined for convenience
\begin{equation}
A={m+1\over m} \left[ {1\over\kappa}\left({2\over\epsilon} +
{1\over z}\right) - {1\over z}\right] - {2m+1\over m\epsilon}.
\end{equation}
Thus, according to Eq.~(21), $G(x)$ is given by
\begin{eqnarray}
G(x)&=&2\xpx H(x) +H(x)+x^m h_m\cr
\noalign{\bigskip}
&=&h_m x^m~+~ {h_m\over Q(x)^2} \left\{ 2m\left(1+Ax\right)x^m~+~
{m(1-x^2)\over Q(x)} \int_0^x dt {t^{m-1}\over Q(t)} \left(1+At\right)
\right\}.
\end{eqnarray}
The properties of $G(x)$ will be discussed in detail in the next
section.

For $D=3$, it is $[(\xpx)^2,x]_-=x(1+2\xpx)$, and one gets for Eq.~(27)
\begin{eqnarray}
&&\left\{Q(x)^2 (\xpx)^2 + Q(x)\left[\xpx Q(x)\right]
\left(1+2\xpx\right)\right\} H(x)\cr
\noalign{\bigskip}
\quad &=& Q(x)\left\{(\xpx)^2 + {\left({1\over\epsilon^2}-1\right)
x^2 \over Q(x)^4}\right\} \left\{Q(x) H(x)\right\}\cr
\noalign{\bigskip}
\quad &=&\left\{\left(1-{x\over\epsilon}\right)m^2 +\left[ {1\over
\kappa} \left({2\over\epsilon}+{1\over z}\right)-\left(
{1\over\epsilon}+{1\over z}\right)\right](m+1)^2 x \right\} x^m h_m.
\end{eqnarray}
For this case, and any other integer $D\geq 3$, there is little hope
to obtain the eigenvalue equation analytically. Local analysis for $H(x)$
is insufficient because the eigenvalue condition requires exact
knowledge of $H(x)$ in the neighborhood of both, the origin {\sl and}
$x=\gamma<1$.

\section{Critical Point Analysis for a Polymer near a Cylindrical
Boundary}

In Eq.~(34), I derived the generating function $G(x)$ for the
eigenvalue problem in Eq.~(13) which describes the behavior of
polymers growing near a cylindrical boundary, $D=2$. It is obvious that
$G(x)$ in general has singularities at $x=\gamma$ and $x=1/\gamma$. If
$\gamma$ is complex, it is $\gamma=1/\gamma^*$,
and both singularities are located on the unit circle. Then it is
$\lambda_{max}=2$, the value of the upper bound in the continuous spectrum
of the transfer matrix $T$, which remains independent of $\kappa$.
Thus, since $z_{\infty}(\kappa)\equiv 1/\lambda_{max}$, it is
$P(\kappa)\propto dz_{\infty}/d\kappa=0$, and the polymer is in the
desorbed phase.

For $P(\kappa)\not=0$, $\kappa$ has to be such that a bound state
$\lambda$ exists which is larger than the continuous spectrum. A bound
state emerges for real positive $\gamma<1$. It can be determined
by imposing the condition in (28). A local analysis of $G(x)$ for
$x\nearrow\gamma$ yields
\begin{equation}
G(x)\sim \left(\gamma-x\right)^{-{3\over 2}} {m h_m\gamma^{m+{3\over2}}
\over \sqrt{1-\gamma^2}} \int_0^1 dt~t^{m-1}(1+A\gamma t) \left[(1-t)
(1-\gamma^2 t)\right]^{-{1\over 2}}~+~ {\rm finite},\quad x\nearrow\gamma.
\end{equation}
Thus, $G(x)$ is finite at $x=\gamma$, if
\begin{equation}
0=\int_0^1 dt~t^{m-1}(1+A\gamma t) \left[(1-t)
(1-\gamma^2 t)\right]^{-{1\over 2}}.
\end{equation}

With $A$ given in Eq.~(33), this relation defines a bound state
$\lambda(\kappa,z)$. For it to exist, it is necessary that
$\gamma<1$, i. e. that $\kappa>\kappa^*(z)$ where $\kappa^*(z)$
is obtained from Eq.~(37) in the (carefully taken) limit
$\gamma\nearrow 1$, $\epsilon\nearrow 1$:
\begin{equation}
\kappa^*(z)={1+2z\over 1+z}.
\end{equation}

In Eq.~(12) I defined the adsorption fraction $P(\kappa)$ in the 
$(\kappa,z)$-plane on the line $z_{\infty}(\kappa)$, where the polymer 
length approaches infinity. The function $z_{\infty}(\kappa)$ is
obtained implicitly from Eq.~(37) for $z=z_{\infty}(\kappa)$ and 
$\lambda=1/z_{\infty}(\kappa)$.

It is now easy to determine the critical point for
the adsorption transition: Since a bound state and, thus, a nonzero 
adsorption fraction first appear for $\gamma=\epsilon=1$, I
obtain from the definition of $\epsilon$ in Eq.~(26), using Eq.~(9),
that $z_{\infty}(\kappa_c)=1/2$. Inserting this value of $z$ into
$\kappa^*(z)$ implies that
\begin{equation}
\kappa_c={4\over 3}
\end{equation}
because the critical point for $P(\kappa)$ is located where the line on
which it is defined,
$z_{\infty}(\kappa)$, intersects with the line $\kappa^*(z)$.
The behavior of $\kappa^*(z)$ and $z_{\infty}(\kappa)$ is summarized
in Fig.~2 for the case $m=0$. Note that for $1<\kappa<\kappa_c$ 
the boundary is already attractive, but not sufficiently attractive to 
adsorb the polymer.

While the value of $P(\kappa)$ for arbitrary $\kappa$ can be obtained 
numerically from the implicit equation for $z_{\infty}(\kappa)$, it
is simple to find the asymptotic behavior near the critical point
explicitly. First, using the integral definition of the
hypergeometric function $F$,\cite{A+S} I rewrite Eq.~(37) as
\begin{eqnarray}
0&=&F_m+\gamma A F_{m+1},\cr
\noalign{\bigskip}
&&F_m={\Gamma({1\over 2})\Gamma(m)\over\Gamma(m+{1\over 2})}
F\left({1\over 2},m;m+{1\over 2};\gamma^2\right).
\end{eqnarray}
If I now substitute
\begin{eqnarray}
z&=&z_{\infty}(\kappa_c)-\Delta z,\quad \Delta z\to 0_+,\cr
\noalign{\medskip}
\kappa&=&\kappa_c+\Delta \kappa,\quad \Delta\kappa\to 0_+,
\end{eqnarray}
and only keep terms to sufficient order, i. e.
\begin{eqnarray}
\epsilon&\sim&1,\cr
\noalign{\medskip}
\gamma&\sim& 1-2\sqrt{3}\Delta z^{1\over 2},\quad\Delta z\to 0_+,
\end{eqnarray}
and use formula 15.3.10 in Ref.~\cite{A+S}
\begin{equation}
F_m\sim -\ln(1-\gamma^2)-\psi(m)+2\psi(1)-\psi({1\over 2}),
\quad \gamma\nearrow 1,
\end{equation}
I get
\begin{equation}
\Delta z\sim {1\over 48} e^{-{8\over 9 (m+1) \Delta\kappa}}.
\end{equation}
With 
\begin{equation}
P(\kappa)\sim {\kappa_c \over z_{\infty}(\kappa_c)} {d\Delta
z\over d\Delta\kappa},
\end{equation}
I finally obtain Eq.~(1). A more extensive calculation shows that
the next-to-leading-order corrections are exponentially smaller yet. 

So far, I have considered the radius $m$ of the attractive boundary to
be a fixed parameter. An interesting crossover phenomenon is revealed
by investigating the asymptotic behavior of the eigenvalue condition in (40) 
when $m$ is allowed to vary with respect to $\Delta\kappa$. It appears that 
for large $m$ the asymptotic relation in Eq.~(1) is invalid when
$\Delta\kappa$ is not sufficiently small compared to $1/m$. Physically,
it is clear that for $m\to\infty$ any $D$-dimensional spherical boundary 
acts like a planar ($D=1$) boundary on the scale of a monomere length.
One would thus expect to find a crossover region, $\Delta\kappa\sim
m^{-\eta}$, for $\Delta\kappa\to 0_+$, $m\gg 1$, and some $\eta>0$, such
that for $\Delta\kappa\ll m^{-\eta}$ a nonlinear relation like Eq.~(1)
is valid while for $1\gg\Delta\kappa\gg m^{-\eta}$ the linear scaling
characteristic of a planar surface is obtained.\footnote{More
complicated scenarios are conceivable but do not seem to be realized.}
This is in fact visible in Fig.~3: While the adsorption fraction for all
$m<\infty$ initially varies exponentially, it eventually has an
intervall of nearly linear behavior that arises the sooner the larger $m$ gets!

I will now prove that for a cylindrical boundary, $D=2$, the crossover
coefficient is $\eta=1$. To that end, I will have to reanalyze the
eigenvalue condition in (40) to find a relation between $\Delta\kappa$
and $\Delta z$ in the limits $\Delta\kappa\to 0_+$ and  $\Delta z\to 0_+$, 
while at the same time $m\to\infty$. Initially, it is not at all clear
whether $1/m$ is smaller than, larger than or of comparable size with
$\Delta\kappa$ and $\Delta z$. It is therefore necessary to make a general
dominant balance analysis.\cite{BeOr} It is convenient to separate the
range of all possibilities into three distinct cases: 
(1) $m\Delta z^{1\over 2}\sim 1$, (2) $m\Delta z^{1\over 2}\ll 1$, 
and (3) $m\Delta z^{1\over 2}\gg 1$, to localize
the region in which the crossover occurs. Clearly, one should
investigate case (1) first: If one finds either
nonlinear or linear behavior there, then one can exclude either case (2) 
or (3) immediately because it would have just the continuation of 
the behavior in case (1).

For case (1), it is simpler to analyze the integral form of the eigenvalue 
condition in (37) which I rewrite by substituting $t=1-(1-\gamma^2)s$ as
\begin{equation}
0=\int_0^{1\over 1-\gamma^2} {ds\over \sqrt{s\left(1+s\right)}}
\left[1-\left(1-\gamma^2\right)s\right]^{m-1} 
\left[1-\left(1-\gamma^2\right){s\over 1+s}\right]^{-{1\over 2}}
\left[1+\gamma A-\gamma A\left(1-\gamma^2\right)s\right].
\end{equation}
Under the conditions for case (1), the integrand is sharply peeked at
$s=0$ and I can use Laplace's method\cite{BeOr} to find the asymptotic
expansion of the integral. Since $1-\gamma^2\sim 4\sqrt{3}\Delta z^{1\over 2}$,
and with 
\begin{equation}
\gamma A\sim 4\left({1\over\kappa_c}-1\right) -
{4\over\kappa_c^2}\Delta\kappa - 
8\sqrt{3}\left({1\over\kappa_c}-1\right)\Delta z^{1\over 2} + 
\left({2\over\kappa_c}-1\right) {1\over m},
\end{equation}
I find that the right-hand side of Eq.~(46) is in leading order asymptotic to
\begin{eqnarray}
&&\int\limits_0^{1\over\delta} {ds\over \sqrt{s\left(1+s\right)}}
e^{m\ln\left(1-4\sqrt{3}\Delta z^{1\over 2}\right)} \left[1+4\sqrt{3}
\Delta z^{1\over 2}\right] 
\left[1+2\sqrt{3}{s\over 1+s}\Delta z^{1\over 2}\right]\cr
\noalign{\medskip}
&&\qquad\left[{4\over\kappa_c}-3 - {4\over\kappa_c^2}\Delta\kappa - 
8\sqrt{3}\left({1\over\kappa_c}-1\right)\left(1+2s\right)\Delta z^{1\over 2} + 
\left({2\over\kappa_c}-1\right) {1\over m}\right],\cr
\noalign{\bigskip}
&&\qquad\qquad m\Delta z^{1\over 2}\gg \delta\gg m\Delta z.
\end{eqnarray}
The conditions on $\delta$ ensure that a sufficiently large integration
interval is kept. They also make $1/\delta$ sufficiently small such
that one can expand the exponential in
Eq.~(48) for all but its leading term, and subsequently integrate to
$s=\infty$ with exponentially small error. Then, to make terms of finite 
order vanish I need to set again $\kappa_c={4\over 3}$. To next order I 
find that
\begin{equation}
\int_0^\infty {ds~e^{-cs}
\over \sqrt{s\left(1+s\right)}} 
\left[{9\over 4}\Delta\kappa-2\sqrt{3}\left(1+c+2s\right)\Delta
z^{1\over 2}\right]
\end{equation}
has to vanish, where $c=4\sqrt{3}m\Delta z^{1\over 2}$ is simply some
number of order unity. Disregarding numbers of order unity, I finally 
get that $\Delta\kappa\sim\Delta z^{1\over 2}$, leading to a linear 
scaling relation by Eq.~(45).
Consequently, the crossover appears to occur already in region (2).

To analyze region (2) I return to Eq.~(40). But to obtain the desired
relation, it is necessary in this case to retain one more term in
formula 15.3.10 in Ref.~\cite{A+S} that was already used in Eq.~(43).
This fact only emerges after trails, each including an increasing number of
terms, and is due to many cancellations among leading-order terms. It is
\begin{eqnarray}
F_m&\sim& -\ln\left(1-\gamma^2\right)-\psi(m)+2\psi(1)-\psi({1\over 2})\cr
\noalign{\medskip}
&&\quad+{1\over 2} m\left(1-\gamma^2\right) \left[-\ln\left(1-\gamma^2\right)-
\psi(m+1)+2\psi(2)-\psi({3\over 2})\right],\cr
\noalign{\bigskip}
&\sim&-\ln\left(4\sqrt{3}m\Delta z^{1\over 2}\right)+2\psi(1)-
\psi({1\over 2}) -2\sqrt{3}m\Delta z^{1\over 2}
\ln\left(4\sqrt{3}m\Delta z^{1\over 2} \right) \cr 
\noalign{\medskip}
&&\quad+ 2\sqrt{3}m\Delta z^{1\over 2}\left[2\psi(2)-\psi({1\over 2})\right]
+{1\over 2m}+ \sqrt{3}\Delta z^{1\over 2}.
\end{eqnarray}
I insert this form into Eq.~(40), expand, and match terms order by order. 
To leading order, I find again that $\kappa_c={4\over 3}$. Thereafter,
many terms cancel until I finally obtain
\begin{equation}
-{9\over 4}\Delta\kappa\ln\left(4\sqrt{3}m\Delta z^{1\over 2}\right)\sim
{1\over m}+2\sqrt{3}\Delta z^{1\over 2}.
\end{equation}
Since in this case $1/m\gg \Delta z^{1\over 2}$, the nonlinear result in
Eq.~(1) is actually obtained for {\sl all} of region (2). Only when the
two terms on the right-hand side of Eq.~(51) balance does Eq.~(1) cease
to hold. But the case $1/m\sim\Delta z^{1\over 2}$ is just beyond the
margin of region (2) and in fact corresponds to region (1).
There, I already found that $1/m\sim\Delta z^{1\over
2}\sim\Delta\kappa$, indicating that $\eta=1$.

\section{Conclusions}

I have shown that the adsorption-desorption transition for a
polymer near an attractive cylindrical boundary is substantially
weaker than in the case of a planar boundary. While the adsorption
fraction vanishes linearly in the planar case for attractive energies
approaching the critical value, it vanishes exponentially fast in the 
cylindrical case.

The nonlinear (i. e. not mean-field like) character of this result 
demonstrates the power of a random walk on hyperspherical lattices 
that was used in this model. Intuitively, it is clear that the exponential 
(inverted logarithmic) behavior observed in this model arises from the
fraction of walks which ``almost'' escape to infinity, i. e. from
walks that would disappear to infinity if the spatial dimension
were to be $D=2+0_+$. This is a simple fact for walks in the critical
dimension $D=2$, independent of the lattice. But while it is not
even obvious how to model this cylindrical geometry onto a more
commonly used lattice like the square lattice, the hyperspherical
lattice makes the model readily solvable. A hypercubic lattice provides
far to much detail about space (which is useful for more
complicated problems). A hyperspherical lattice on the other hand
reduces any $D$-dimensional space to a simple one-dimensional
configuration which provides sufficient detail for spherically symmetric
problems in statistical mechanics such as this. Thus, it can be expected
that this lattice will lead to many more insights into the critical
behavior of simple problems that otherwise are intractable on
hypercubic lattices.

In a future paper,\cite{BeBoMe} with further simplifications in
the walk model on hyperspherical lattices,\cite{BeCoMe} it will
be argued that the adsorption fraction scales with a critical
coefficient $(2-|2-D|)/|2-D|$ for $1\leq D<4$, $D\not=2$, while it
shows a first order transition for $D>4$.

\section*{Acknowledgements}

I thank especially Moshe Moshe for inviting me to the Technion and 
for discussing many aspects of this problem with me. I much appreciated 
the hospitality at the Technion.  Further, I would like to
thank Carl Bender and Peter Meisinger for
useful discussions. This work was supported by 
the  U.S. Department of Energy under Contract No. DE-AC02-76-CH00016.

\newpage
\section*{FIGURE CAPTIONS}

\noindent
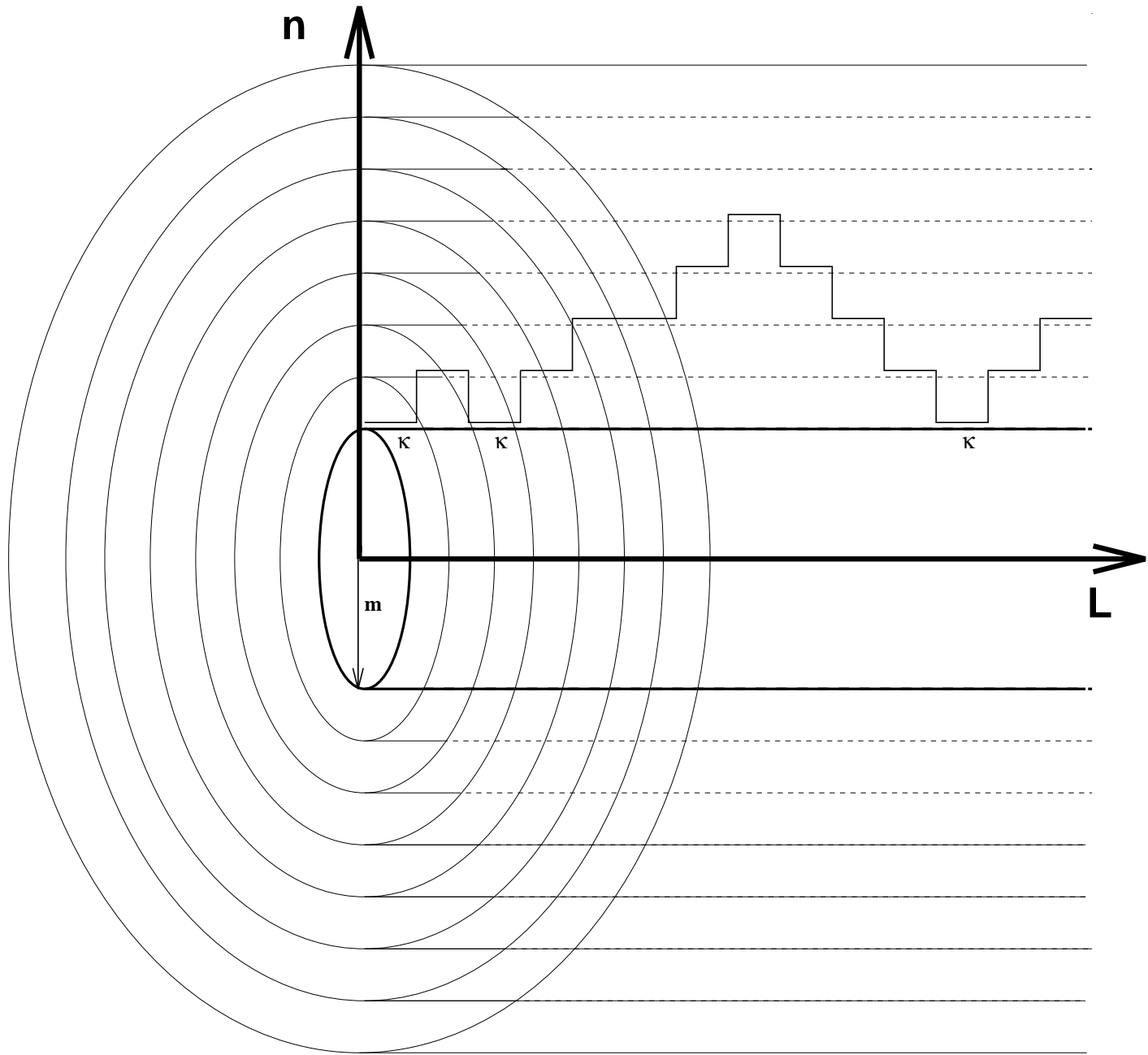
FIGURE 1: Random walk on a lattice consisting of concentric cylindrical 
surfaces of unit radii. Such a walk serves as model for a polymer growing 
at an attractive cylindrical boundary with radius $m$ (thickened
lines). The polymer is initially grafted to the boundary and is growing
to the right. Every time a monomer gets added at the boundary, the
polymer gains in potential energy by an amount $\kappa$.

\noindent
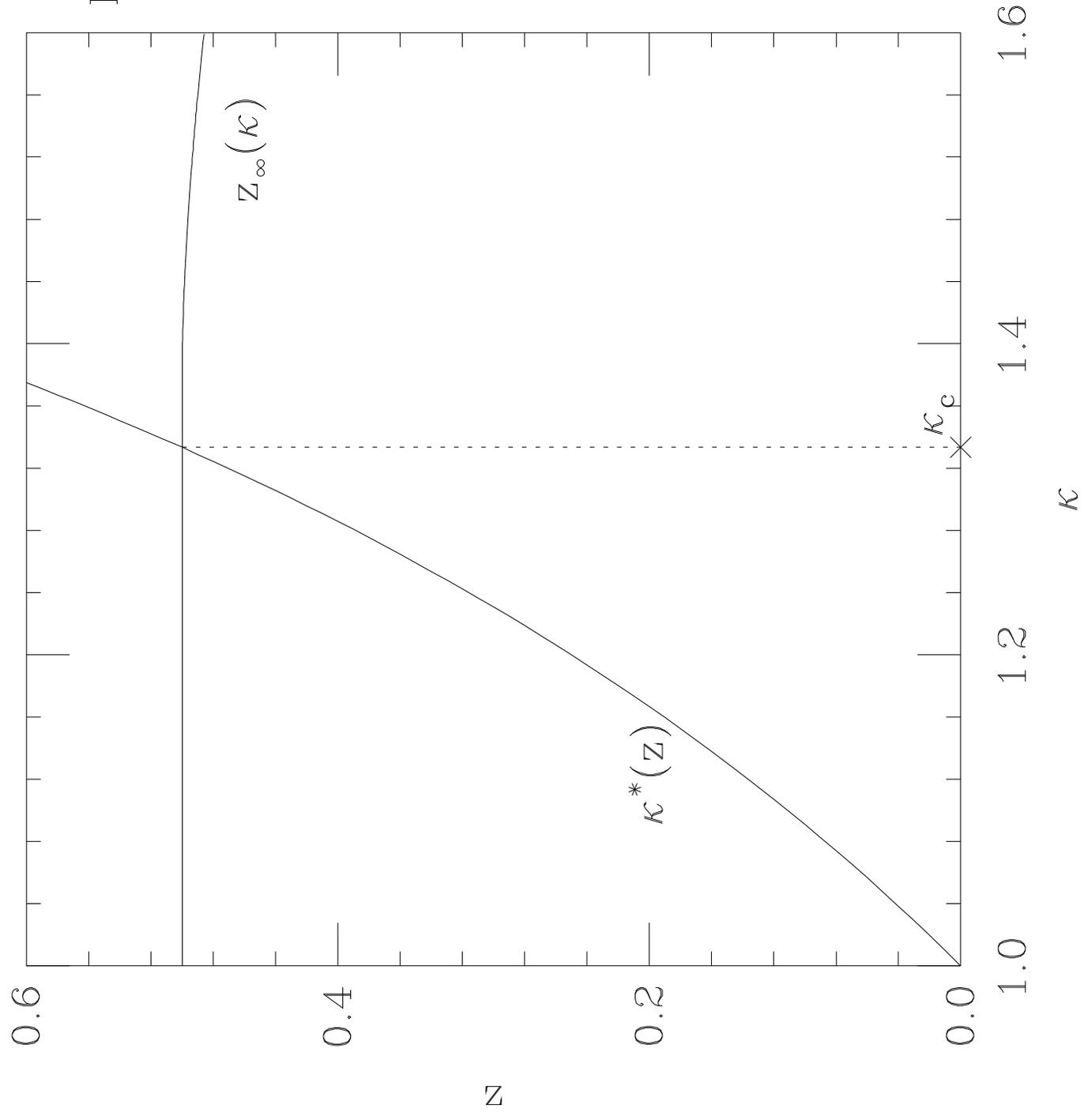
FIGURE 2: The $(\kappa,z)$-plane with the lines $\kappa^*(z)$ and
$z_{\infty}(\kappa)$ for $m=0$. The adsorption fraction $P(\kappa)$
is defined only on the line $z_{\infty}(\kappa)$, where the polymer 
reaches infinite length. An adsorbed phase, $P(\kappa)\not=0$, only 
exists when $\kappa>\kappa^*(z)$. Thus, the critical point 
$\kappa_c$ is given by the value of $\kappa$ at the intersection of 
both curves.

\noindent
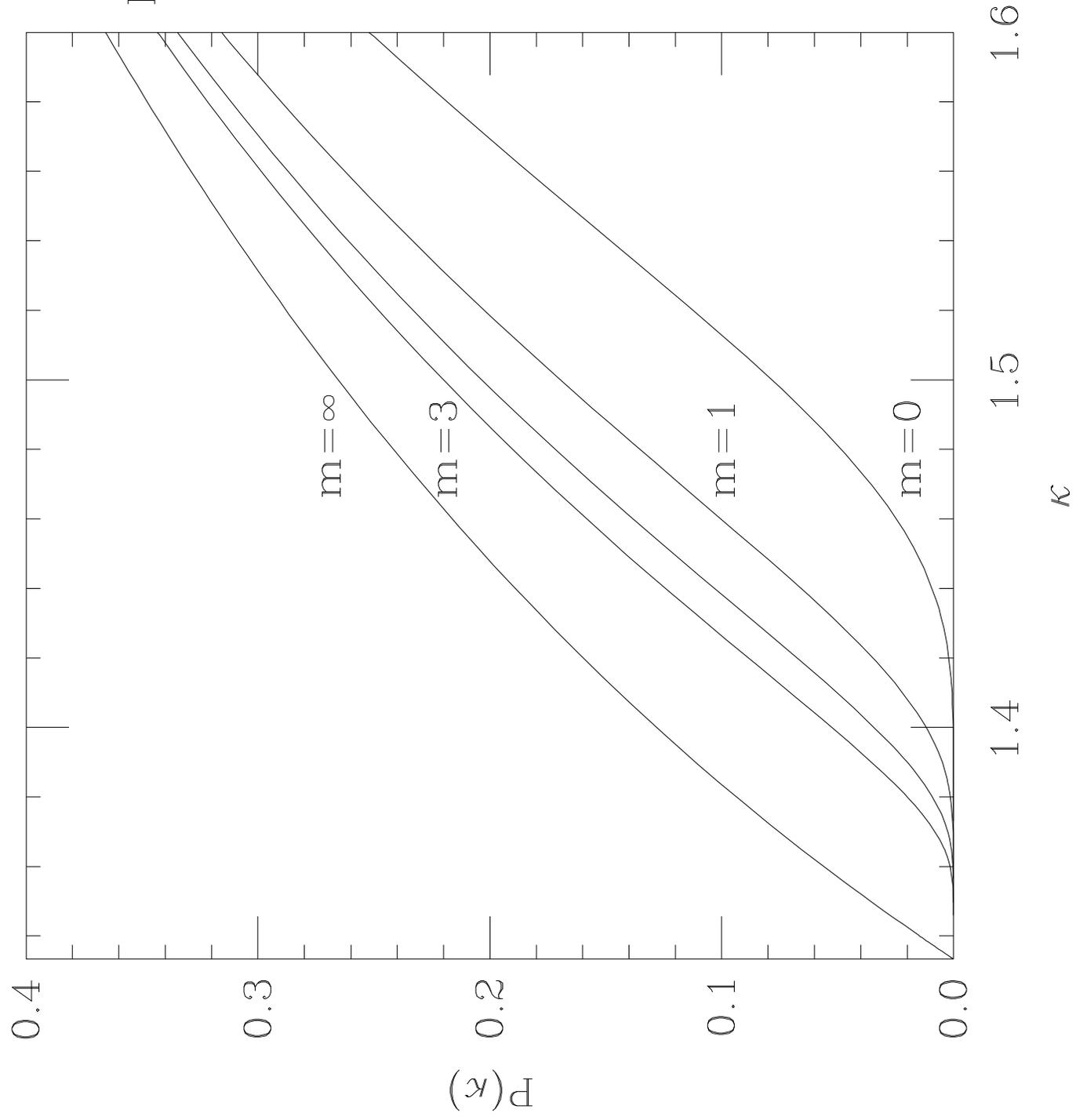
FIGURE 3: The exact adsorption fraction $P(\kappa)$ plotted for 
$m=0,~1,~2,~3$ and $m=\infty$ and $\kappa\geq\kappa_c=4/3$.
{}For finite $m$, $P(\kappa)$ vanishes
exponentially for $\kappa\searrow\kappa_c$ while it crosses over
to a linear  increase when $\kappa-\kappa_c\sim 1/m$. For
$m=\infty$ I recover the linear behavior for $P(\kappa)$ for 
$\kappa\searrow\kappa_c$ that was found  in Ref.~\cite{Priv}.


\newpage
\begin{thebibliography} {99}
\bibitem{AIP} 
See, for example, {\it Random Walks and Their Application  in the
Physical and Biological Sciences}, eds. M. F. Shlesinger  and B.
J. West (American Institute of Physics, New York, 1984).
\bibitem{DesCJ} 
J. Des Cloizeaux and G. Jannink, {\it Polymers in Solution},
(Clarendon, Oxford, 1990).
\bibitem{DeGennes} 
P. G. DeGennes, {\it Scaling Concepts in Polymer Physics}, (Cornell,
New York, 1979).
\bibitem{Freed} 
K. F. Freed, {\it Renormalization Group Theory of Macromolecules},
(Wiley, New York, 1987).
\bibitem{Binder}
K. Binder and K. Kremer, in {\it Scaling Phenomena in Disordered
Systems}, eds. R. Rynn and A. Skjeltorp (Plenum, New York, 1985),
K. Binder, Adv. Polymer Sci. {\bf 112} 181 (1994), and K. Binder, Nucl.
Phys. B, Proceeding Supplement for the LATTICE'94 Conference in
Bielefeld, Germany, (to appear).
\bibitem{Baxter} 
R. J. Baxter, {\it Exactly Solved Models in
Statistical Mechanics}, (Academic, London, 1982).
\bibitem{FloMo} 
P. J. Flory, J. Chem. Phys. {\bf 17}, 303 (1949); E. W. Montroll,
J. Chem. Phys. {\bf 18}, 734 (1950).
\bibitem{Sokal}
R. Fernandez, J. Fr\"ohlich, and A. D. Sokal, {\it Random Walks,
Critical Phenomena, and Triviality in Quantum Field Theory},
(Springer, Berlin, 1992).
\bibitem{Priv}
V. Privman, G. Forgacs, and H. L. Frisch, Phys. Rev. B {\bf 37},
9897 (1988), and V. Privman and N. M. \v Svraki\'c, {\it Directed
Models of Polymers, Interfaces, and Clusters: Scaling and Finite-Size 
Properties}, (Springer, Berlin, 1989).
\bibitem{BeBoM}
C. M. Bender, S. Boettcher, and L. R. Mead, J. Math. Phys. {\bf 35},
368 (1994), and C. M. Bender, S. Boettcher, and M. Moshe, J. Math.
Phys. {\bf 35}, 4941  (1994).
\bibitem{BoMo}
The results of this calculation were presented in brief in 
S. Boettcher and M. Moshe, {\it ``Statistical Models on Spherical
Geometries''} (submitted).
\bibitem{PliBer}
See, for instance, M. Plischke and B. Bergersen, {\it Equilibrium
Statistical Mechanics}, pp. 163, (Prentice Hall, New Jersey, 1989).
\bibitem{Taka}
G. H. Weiss and R. J. Rubin, Adv. Chemical Physics {\bf 52}, 363
(1983); A. Takahashi and M. Kawaguchi, Adv. Polymer Sci. {\bf 46},
1 (1982); Yu. S. Lipatov and L. M. Sergeeva, {\it Adsorption of
Polymers}, (Halsted, Jerusalem, 1974).
\bibitem{A+S}
{}F. Oberhettinger, in {\it Handbook of Mathematical Functions}, eds.
M. Abramowitz and I. Stegun, pp. 555, (National Bureau of Standards, 
Washington, 1972).
\bibitem{BeOr}
C. M. Bender and S. A. Orszag, {\it Advanced Mathematical Methods for
Scientists and Engineers}, (McGraw-Hill, New York, 1978).
\bibitem{BeBoMe}
C. M. Bender, S. Boettcher, and P. N. Meisinger, in preparation.
\bibitem{BeCoMe}
C. M. Bender, F. Cooper, and P. N. Meisinger, in preparation.

\end{thebibliography}
\end{document}